\documentclass[english]{acmart}

\usepackage[T1]{fontenc}
\usepackage[latin9]{inputenc}
\usepackage{babel}
\usepackage{array}
\usepackage{booktabs}
\usepackage{textcomp}
\usepackage{multirow}
\usepackage{graphicx}
\usepackage{wasysym}
\ifx\hypersetup\undefined
  \AtBeginDocument{%
    \hypersetup{unicode=true,
 bookmarks=false,
 breaklinks=false,pdfborder={0 0 1},backref=false,colorlinks=false,
 pdfborderstyle=,nolinks=true}
  }
\else
  \hypersetup{unicode=true,
 bookmarks=false,
 breaklinks=false,pdfborder={0 0 1},backref=false,colorlinks=false,
 pdfborderstyle=,nolinks=true}
\fi

\makeatletter

\newcommand{\lyxmathsym}[1]{\ifmmode\begingroup\def\b@ld{bold}
  \text{\ifx\math@version\b@ld\bfseries\fi#1}\endgroup\else#1\fi}

\providecommand{\tabularnewline}{\\}
\floatstyle{ruled}
\newfloat{algorithm}{tbp}{loa}
\providecommand{\algorithmname}{Algorithm}
\floatname{algorithm}{\protect\algorithmname}

\usepackage{babel}
\makeatother

\begin{document}
\begin{abstract}
Polynomial multiplication stands out as a highly demanding arithmetic
process in the development of post-quantum cryptosystems. The importance
of the number-theoretic transform (NTT) extends beyond post-quantum
cryptosystems, proving valuable in enhancing existing security protocols
such as digital signature schemes and hash functions. CRYSTALS-KYBER
stands out as the sole public key encryption (PKE) algorithm chosen
by the National Institute of Standards and Technology (NIST) in its
third round selection, making it highly regarded as a leading post-quantum
cryptography (PQC) solution. Due to the potential for errors to significantly
disrupt the operation of secure, cryptographically-protected systems,
compromising data integrity, and safeguarding against side-channel
attacks initiated through faults it is essential to incorporate mitigating
error detection schemes. This paper introduces algorithm level fault
detection schemes in the NTT multiplication using Negative Wrapped
Convolution and the NTT tailored for Kyber Round 3, representing a
significant enhancement compared to previous research. We evaluate
this through the simulation of a fault model, ensuring that the conducted
assessments accurately mirror the obtained results. Consequently,
we attain a notably comprehensive coverage of errors. Furthermore,
we assess the performance of our efficient error detection scheme
for Negative Wrapped Convolution on FPGAs to showcase its implementation
and resource requirements. Through implementation of our error detection
approach on Xilinx/AMD Zynq Ultrascale+ and Artix-7, we achieve a
comparable throughput with just a 9\% increase in area and 13\% increase
in latency compared to the original hardware implementations. Finally,
we attained an error detection ratio of nearly 100\% for the NTT operation
in Kyber Round 3, with a clock cycle overhead of 16\% on the Cortex-A72
processor.
\end{abstract}

\title{Efficient Algorithm Level Error Detection for Number-Theoretic Transform
used for Kyber Assessed on FPGAs and ARM}
\author{Kasra Ahmadi, Saeed Aghapour, Mehran Mozaffari Kermani\textit{, }and
Reza Azarderakhsh}
\authornote{K. Ahmadi, S. Aghapour, and M. Mozaffari-Kermani are with the Department
of Computer Science and Engineering, University of South Florida,
Tampa, FL 33620, USA. e-mails: \{ahmadi1, aghapour, mehran2\}@usf.edu.\\
R. Azarderakhsh is with the Department of Computer and Electrical
Engineering and Computer Science, Florida Atlantic University, Boca
Raton, FL 33431, USA. e-mail: razarderakhsh@fau.edu.}
\maketitle

\section{Introduction}

Fast Fourier Transform (FFT) \citep{cochran1967fast} algorithms which
are used to compute the Discrete Fourier Transform (DFT) have various
applications, ranging from digital signal processing to the efficient
multiplication of large integers. When the coefficients of the polynomial
are specifically chosen from a finite field, the resulting transform
is known as the Number Theoretic Transform (NTT) \-\citep{cooley1965algorithm},
and it can be computed using FFT algorithms designed for operations
within this particular finite field. An efficient approach for polynomial
multiplication, NTT holds significant importance in post-quantum cryptosystems,
e.g., lattice-based cryptosystems which are regarded as a leading
contender for quantum-secure public key cryptography, primarily because
of its broad applicability and security proofs grounded in the worst-case
hardness of established lattice problems. 

In 2022, NIST unveiled the Round 3 outcomes of the Post-Quantum Cryptography
standardization Process, which featured four chosen algorithms (CRYSTALS-KYBER
, CRYSTALS-DILITHIUM, Falcon, SPHINCS+). These are called formally
after Aug. 2024 with respect to Federal Information Processing Standards
(FIPS) as FIPS 203, Module-Lattice-Based Key-Encapsulation Mechanism
Standard, FIPS 204, Module-Lattice-Based Digital Signature Standard,
and FIPS 205, Stateless Hash-Based Digital Signature Standard while
Falcon is yet to get a FIPS number \citep{FIPS203}, \citep{FIPS204},
and \citep{FIPS205}. Out of the chosen algorithms, CRYSTALS-KYBER,
also referred to as Kyber \citep{bos2018crystals} and \citep{FIPS203},
stands out as the sole algorithm for both public key encryption and
key establishment that relies on the challenge of the module learning
with errors (MLWE) problem. 

The NTT has proven to be a potent tool that facilitates the computation
of this operation with quasi-linear complexity $O(n.lgn)$. Several
recent studies have been conducted on optimizing the NTT \citep{bisheh2021high},
\citep{geng2023rethinking},\citep{hermelink2023adapting}, \citep{hwang2022verified},
\citep{kurniawan2023configurable}, \citep{li2023scalable}, \citep{mu2023energy},
\citep{tosun2024zero}, and \citep{zhang2021towards}. In addition
to polynomial multiplication, the use of the NTT has the potential
to significantly enhance existing schemes by improving their security
parameters. NTT is widely used in signature schemes, hash functions,
and identification schemes. As a result, incorporating efficient error
detection mechanisms in the NTT for polynomial multiplication will
not only improve the security and reliability but also mitigate the
risk of fault attacks in the respective algorithms in post-quantum
cryptography. Moreover, such schemes could at least alleviate the
attack surface or be add-ons to other approaches. Kyber and Dilithium
employ polynomial multiplication over $\mathbb{Z}[X]/(X^{n}+1)$ utilizing
the NTT. Integrating effective error detection schemes into the NTT
used in Kyber will enhance security, reliability, and mitigate the
risk of fault attacks in post-quantum cryptography algorithms.

\subsection{Related works}

Several previous studies have concentrated on the implementation and
fault detection in different arithmetic components of both classical
and post-quantum cryptography \citep{canto2022reliable}, \citep{guo2024eclbc},
\citep{libano2023efficient}, \citep{richter2020concurrent}, and
\citep{sarker2022efficient}. We categorize relevant studies into
three subsets. The initial category centers on fault attacks and countermeasures
applied to post-quantum cryptographic schemes, particularly those
extensively utilizing the NTT. Numerous prior studies have explored
fault attacks with the objective of gaining access to or compromising
secret keys \citep{elghamrawy2023mlwe}, \citep{heinz2022combined},
\citep{hermelink2021fault}, \citep{ravi2024side}, \citep{saha2023learn},
and \citep{xagawa2021fault}. Ravi \emph{et al.} \citep{ravi2023fiddling}
presented the first fault injection analysis of the NTT. The authors
found a significant flaw in the way the NTT is implemented in the
pqm4 library \citep{kannwischer2019pqm4}. The identified vulnerability,
referred to as twiddle-pointer, is exploited to demonstrate practical
and effective attacks on Kyber and Dilithium \citep{avanzi2019crystals}.
In \citep{ravi2019exploiting}, the authors proposed a fault attack
aiming the NTT operation during the key generation and encapsulation
routine of the Kyber. The fault attack is achieved by zeroize all
the twiddle factors used in the NTT operation. When focusing on applying
this technique to the NTT regarding secrets or errors, it can significantly
diminish the randomness of the secret or error and reduce the entropy.
Several other authors have reported employing fault injection on structured
lattice-based schemes as a foundation for attacks \citep{bindel2016lattice},
\citep{bruinderink2018differential}, and \citep{mus2020quantumhammer}
threatening the security of implementations. Regarding active attacks,
although some prior research addressed fault injection, Espitau et
al. \citep{espitau2018loop} introduced loop-abort faults in several
lattice-based cryptosystems, including CRYSTALS-Kyber. In this attack,
a fault is injected into the cryptosystem, causing the loop responsible
for sampling random Gaussian secret coefficients to terminate early.
This early termination leads to the generation of lower-dimensional
secrets, which can then be leveraged for a key recovery attack. In
2021, Pessl and Prokop \citep{prokop2021fault} proposed an attack
that involves a single instruction-skipping fault during the decoding
process. Their fault simulations showed that at least 6,500 faulty
decapsulation attempts are needed to fully recover the key for Kyber512
operating on a Cortex M4. In the same year, Hermelink et al. \citep{hermelink2021fault}
combined fault injections with chosen-ciphertext attacks on CRYSTALS-Kyber,
suggesting that their attack could bypass defenses such as decoder
shuffling. Their findings demonstrated successful secret key recovery
using 7,500 inequalities for Kyber-512, 10,500 for Kyber-768, and
11,000 for Kyber-1024. Delvaux \citep{delvaux2021roulette} refined
the side-channel attack (SCA) from \citep{hermelink2021fault}, making
it simpler to execute and more challenging to defend against. In \citep{kundu2024carry},
the authors proposed a new fault attack on the SCA-secure masked decapsulation
algorithm for generic LWE-based KEMs and detailed the attack for Kyber.
Jendral \citep{jendral2024single} presented an attack on CRYSTALS-Dilithium
implementation on ARM Cortex-M4 using fault injection. Krahmer \emph{et
al. }\citep{krahmer2024correction} presented two key-recovery fault
attacks on Dilithium's signing procedure.

The focus of the second category lies in offering fault detection
methodologies pertaining to the algorithmic level of classical cryptographic
schemes. In \citep{ahmadi2023efficient}, the authors suggested an
effective algorithmic-level error detection for the ECSM window method,
aiming to identify both permanent and transient errors. In \citep{10587011},
the authors presented algorithm level error detection scheme designed
for Montgomery ladder ECSM algorithm utilized in non-supersingular
elliptic curves.

The third category focuses on developing fault detection schemes specifically
for the NTT. In \citep{bauer2024fault}, the authors introduced a
method for safeguarding the NTT against fault attacks. Mishra \emph{et
al}. \citep{mishra2024probabilistic} introduced a principle for countermeasures
that is based on cryptographic guarantees rather than relying on ad
hoc methods, aiming to offer measurable protection against the previously
mentioned fault attacks on lattice-based schemes. In \citep{singh2024analysis},
the authors integrate the advantages of bit slicing, a software implementation
technique where a datapath of an $n$-bit processor is treated as
$n$ parallel single-bit datapaths, to devise a fault countermeasure
for the NTT used in Dilithium \citep{ducas2018crystals}. Sarker \emph{et
al}. \citep{sarker2022error} and \citep{sarker2018hardware}, presented
error detection architectures of the NTT based on recomputation with
encoded operands. The authors achieved high error coverage with low
area overhead. However, due to the recomputation process, their latency
is doubled. We note that despite high error coverage, for fast implementations,
these works might increase the total time to levels not acceptable.
Additionally, Cintas-Canto et al. \citep{canto2022error} introduced
error detection schemes for lattice-based KEMs using recomputation
techniques and implemented these schemes on FPGA. Below, we present
our major contributions in this work.

\subsection{Our Major Contributions}
\begin{itemize}
\item Our proposed error detection schemes donot depend on recomputation;
instead, it relies on algorithm-level coherency. As a result, it offers
higher speed, lower latency, and reduced area compared to previous
approaches.
\item We have proposed an algorithm level fault detection scheme of the
NTT multiplication using Negative Wrapped Convolution which is widely
used in lattice-based cryptography. We achieved high error coverage
with less area overhead and latency compared to previous works. This
was achieved through the implementation of algorithm-level error detection
for the NTT section and partial recomputation for pre-computation
section. 
\item As the Negative Wrapped Convolution method does not work for the NTT
multiplication used in Kyber Round 3, we have introduced an algorithm
level error detection scheme for the NTT multiplication tailored for
Kyber Round 3. Implementing algorithm level error detection for the
NTT module enabled us to achieve substantial error coverage with minimal
overhead compared to previous works. This method can be integrated
into Kyber Round 3 reference implementation.
\item We performed simulations for single and burst fault injection using
our proposed schemes. The simulation outcomes demonstrated that our
approach is capable of detecting diverse types of faults with high
error coverage and offers protection against the NTT fault attack
presented in \citep{ravi2023fiddling} and those with respective fault
models.
\item Our error detection method for Negative Wrapped Convolution was deployed
on Xilinx/AMD Zynq Ultrascale+ , and Artix-7. The results of our implementations
indicate that we can attain a significantly high level of error coverage
with only a 13\% increase in latency and a 9\% area bloat. Our proposed
error detection method for Kyber Round 3 is implemented on Intel Core-i7,
and Cortex-A72 with additional overhead of 28\% and 16\% in terms
of clock cycle overhead.
\end{itemize}
Our work is structured as follows: Section 2 discusses Negative Wrapped
Convolution and the NTT used in Kyber. The presented error detection
schemes on Negative Wrapped Convolution and the NTT tailored for Kyber
are described in Section 3. Section 4 is divided into two subsections.
First, the fault model in this work is discussed; next, the fault
simulation is performed to assess error detection rate. We apply such
fault detection schemes into the reference Kyber implementation and
Negative Wrapped Convolution in Section 5 to benchmark the different
overheads. Finally, our conclusions are presented in Section 6.

\section{Preliminaries}

\subsection{Number Theoretic Transform (NTT)}

Multiplication of two polynomials $f$ and $g$ take quadratic complexity
$O(n^{2})$ utilizing school book algorithm. However, by using the
NTT, it can be reduced to quasi-linear complexity $O(n.lgn)$. The
NTT represents a specialized version of the Discrete Fourier Transform
(DFT), utilizing a coefficient ring chosen from a finite field that
encompasses the necessary roots of unity. We denote the ring $\mathbb{Z}[X]/(X^{n}+1)$
by $R$ and the ring $\mathbb{Z}_{q}[X]/(X^{n}+1)$ by $R_{q}.$ Consider
$\omega$ to be a primitive $n$-th root of unity in $\mathbb{Z}_{q}$
which $\omega^{n}\,\equiv\,1$ (mod $q$) where $q\,\equiv\,1\,$(mod
$2n$), $q$ is a prime number, and $n$ is power of 2. The NTT of
a polynomial $f\in R_{q}$ is defined as $\hat{f}=\text{NTT}(f)=\sum_{j=0}^{n-1}f[j]\omega^{ij}\,$(mod
$q$) and the inverse NTT $f=\text{NTT}^{-1}(\hat{f})=\sum_{j=0}^{n-1}n^{-1}\hat{f}[j]\omega^{-ij}\,(\text{mod }q)$.
The NTT of a sequence $f$ is derived as in the form of matrix multiplication
described in (1). The Cooley-Tukey butterfly algorithm can be used
to efficiently implement the forward NTT.

\begin{equation}
\hat{f}=\text{NTT(}f)=\left[\begin{array}{cccc}
\omega^{0} & \omega^{0} & ... & \omega^{0}\\
\omega^{0} & \omega^{1} & ... & \omega^{n-1}\\
\omega^{0} & \omega^{2} & ... & \omega^{2(n-1)}\\
. & . & ... & .\\
\omega^{0} & \omega^{n-1} & ... & \omega^{(n-1)^{2}}
\end{array}\right]*\left[\begin{array}{c}
f(0)\\
f(1)\\
f(2)\\
.\\
f(n-1)
\end{array}\right].
\end{equation}

In Algorithm 1, the iterative NTT implementation derives the NTT of
a specified polynomial $f$. The $\text{bit\_reverse}(k)$ function
(presented in Line 5) rearranges the input $k$ where the new placement
of elements is determined by reversing the binary representation of
the operand. Lines 9 and 10 perform the butterfly operation. Every
butterfly module, taking inputs $a$ and $b$ and producing outputs
$c$ and $d$, executes the following butterfly computation: $c=a+b\,\omega^{k}$
and $d=a-b\,\omega^{k}$. Kyber's reference implementation uses Montgomery
multiplication to achieve efficient modular multiplication and improve
performance by converting numbers into Montgomery form.

\begin{algorithm}[t]
\begin{raggedright}
\caption{{\large{}Iterative NTT \citep{Zhang2019}}}
\textbf{\large{}Input: $f\in\mathbb{Z}_{q}$}{\large{} of length $n=2^{k}$,
$\omega$ is primitive $n$-th root of unity}{\large\par}
\par\end{raggedright}
\begin{raggedright}
\textbf{\large{}Output: }{\large{}$\hat{f}=\text{NTT}(f)$ in bit-reversed
order}{\large\par}
\par\end{raggedright}
\begin{raggedright}
1: $\hat{f}=f$
\par\end{raggedright}
\begin{raggedright}
2: \textbf{for} $s=k$ \textbf{to} $1$:
\par\end{raggedright}
\begin{raggedright}
3: $\qquad$$m=2^{s}$
\par\end{raggedright}
\begin{raggedright}
4: $\qquad$\textbf{for} $k=0$ \textbf{to} $2^{k-s}-1$:
\par\end{raggedright}
\begin{raggedright}
5:\textbf{ }$\qquad$$\qquad$ \textbf{$\omega=\omega^{\text{bit-reverse}(k)\,.\,m/2}$}
\par\end{raggedright}
\begin{raggedright}
6: $\qquad\qquad$ \textbf{for $j=0$ to $m/2-1$}:
\par\end{raggedright}
\begin{raggedright}
7: $\qquad\qquad\qquad$ $u=\hat{f}[k.m+j]$
\par\end{raggedright}
\begin{raggedright}
8: $\qquad\qquad$$\qquad$ $t\equiv\omega\,.\,\hat{f}[k.m+j+m/2]$
mod $q$
\par\end{raggedright}
\begin{raggedright}
9: $\qquad\qquad$$\qquad$ $\hat{f}[k.m+j]\equiv(u+t)$ mod $q$
\par\end{raggedright}
\begin{raggedright}
10:$\qquad\qquad$$\qquad$$\hat{f}[k.m+j+m/2]\equiv(u-t)$ mod $q$
\par\end{raggedright}
\raggedright{}11: \textbf{return} $\hat{f}$
\end{algorithm}

\subsection{Negative Wrapped Convolution}

Let $f,g\in$$R_{q}$. Computing $h=f\,\text{.}\,g$ mod $x^{n}+1$
requires applying the NTT of length $2n$ and $n$ zeros to be extended
to $f$ and $g$. This essentially doubles the input length and also
necessitates an explicit reduction modulo $X^{n}+1$ . This problem
can be resolved by utilizing Negative Wrapped Convolution, which prevents
the doubling of input length \citep{lyubashevsky2008swifft}. Let
$\psi$ be a primitive 2$n$-th root of unity in $\mathbb{Z}_{q}$
where $\psi^{2}=\omega$. 

Moreover, there exists a Pre-process module which is described in
Algorithm 2. The Negative Wrapped Convolution of $f$ and $g$ is
defined as $h=\text{post-process}(\text{INTT}(\text{NTT}(\tilde{f})\,\Circle\text{\,NTT}(\tilde{g})))$
where $\tilde{f}=\text{pre-process}(f)$, $\tilde{g}=\text{pre-process}(g)$,
and the symbol $\Circle$ represents component-wise multiplication.
We can achieve Post-process function by changing $\psi$ to $\psi^{-1}$.
Post-process function is described in Algorithm 3.
\begin{algorithm}[t]
\begin{raggedright}
\caption{{\large{}The pre-process step for the NTT calculations}}
\par\end{raggedright}
\begin{raggedright}
\textbf{\large{}Input: $f\in\mathbb{Z}_{q}$}{\large{} of length $n$,
$\psi$ primitive 2$n$-th root of unity}{\large\par}
\par\end{raggedright}
\begin{raggedright}
\textbf{\large{}Output: }{\large{}$\tilde{f}=\text{pre-process}(f)$}{\large\par}
\par\end{raggedright}
\begin{raggedright}
1: \textbf{for} $i=0$ \textbf{to} $n-1$:
\par\end{raggedright}
\begin{raggedright}
2: $\qquad$$\tilde{f}[i]=f[i]\,.\,$$\psi^{i}$
\par\end{raggedright}
\raggedright{}3:\textbf{ return} $\tilde{f}$
\end{algorithm}
\begin{algorithm}[t]
\begin{raggedright}
\caption{{\large{}The Post-process step for the NTT calculations}}
\par\end{raggedright}
\begin{raggedright}
\textbf{\large{}Input: $f\in\mathbb{Z}_{q}$}{\large{} of length $n$,
$\psi$ primitive 2$n$-th root of unity}{\large\par}
\par\end{raggedright}
\begin{raggedright}
\textbf{\large{}Output: }{\large{}$\tilde{f}=\text{post-process}(f)$}{\large\par}
\par\end{raggedright}
\begin{raggedright}
1: \textbf{for} $i=0$ \textbf{to} $n-1$:
\par\end{raggedright}
\begin{raggedright}
2: $\qquad$$\tilde{f}[i]=f[i]\,.\,$$\psi^{-i}$
\par\end{raggedright}
\raggedright{}3:\textbf{ return} $\tilde{f}$
\end{algorithm}
\begin{algorithm}[t]
\begin{raggedright}
\caption{{\large{}Kyber NTT}}
\textbf{\large{}Input: $r\in\mathbb{Z}_{q}$}{\large{} of length $n=256$,
$\omega=17$ is primitive $n$-th root of unity}{\large\par}
\par\end{raggedright}
\begin{raggedright}
\textbf{\large{}Output: }{\large{}$\text{NTT}(r)$ in bit-reversed
order}{\large\par}
\par\end{raggedright}
\begin{raggedright}
1: $\widetilde{r}=r$
\par\end{raggedright}
\begin{raggedright}
2: $j=0$
\par\end{raggedright}
\begin{raggedright}
3: $k=0$
\par\end{raggedright}
\begin{raggedright}
4: \textbf{for (}$s=128;s>=2;s=s/2$):
\par\end{raggedright}
\begin{raggedright}
5: $\qquad$\textbf{for} ($i=0;i<256;i=j+s$):
\par\end{raggedright}
\begin{raggedright}
6:\textbf{ }$\qquad$$\qquad$ \textbf{$zeta=\omega^{\text{(bit-reverse}(k))}$}
\par\end{raggedright}
\begin{raggedright}
7:\textbf{ }$\qquad$$\qquad$ \textbf{$k=k+1$}
\par\end{raggedright}
\begin{raggedright}
8: $\qquad\qquad$ \textbf{for} $(j=i;j<i+s;j++)$
\par\end{raggedright}
\begin{raggedright}
9: $\qquad\qquad\qquad$ $t=\text{fqmul}(zeta,\widetilde{r}[j+s])$
\par\end{raggedright}
\begin{raggedright}
10:$\qquad\qquad$$\qquad$ $\widetilde{r}[j+s]=\widetilde{r}[j]-t$
mod $q$
\par\end{raggedright}
\begin{raggedright}
11:$\qquad\qquad$$\qquad$$\widetilde{r}[j]=\widetilde{r}[j]+t$
mod $q$
\par\end{raggedright}
\raggedright{}12: \textbf{return} $\widetilde{r}$
\end{algorithm}

\subsection{Polynomial Multiplication Utilizing the NTT in Kyber}

To use Negative Wrapped Convolution method, we require that $2n\mid(q-1)$.
Regarding the parameters used in Kyber with the prime $q=3329$ and
$n=256$ the base field $\mathbb{Z}_{q}$ contains 256-th roots of
unity but not 512-th roots. Therefore, we cannot use the method used
in \citep{lyubashevsky2008swifft}. To overcome this problem, the
polynomial $X^{256}+1$ of $R$ factor into 128 polynomials of degree
2 modulo $q$. The polynomial $X^{256}+1$ can be written as
\[
X^{256}+1=\prod_{i=0}^{127}(X^{2}-\omega^{2i+1})=\prod_{i=0}^{127}(X^{2}-\omega^{2b_{7}(i)+1}),
\]
where $b_{7}(i)$ is the bit reversal of the 7-bit $i$. Therefore,
the NTT of $f$ is a vector of 128 polynomials of degree one.
\[
\hat{f}=\text{NTT(}f)=(\hat{f}_{0}+\hat{f}_{1}X,\hat{f}_{2}+\hat{f}_{3}X,...,\hat{f}_{254}+\hat{f}_{255}X),
\]
with

\begin{equation}
\hat{f}_{2i}=\sum_{j=0}^{127}f_{2j}\omega^{(2br_{7}(i)+1)j},
\end{equation}
\begin{equation}
\hat{f}_{2i+1}=\sum_{j=0}^{127}f_{2j+1}\omega^{(2br_{7}(i)+1)j}.
\end{equation}

We can compute $h=\hat{f}\,\lyxmathsym{\textopenbullet}\,\hat{g}=\text{NTT}^{-1}(\text{NTT}($$f$)$\,$\textopenbullet $\,\text{NTT(}g$))
consisting of the 128 products of linear polynomials
\begin{align}
\hat{h}_{2i}+\hat{h}_{2i+1}X & =(\hat{f}_{2i}+\hat{f}_{2i+1}X)(\hat{g}_{2i}+\hat{g}_{2i+1}X)\text{ mod (\ensuremath{X^{2}}-\ensuremath{\omega^{2b_{7}(i)+1}})}.
\end{align}

The iterative NTT implementation of Kyber is described in Algorithm
4. fqmul function (presented in Line 9 of Algorithm 4) performs multiplication
of two operand and Montgomery reduction afterward.

\section{Proposed Error Detection Scheme}

\subsection{Negative Wrapped Convolution}

In this section, we present our error detection schemes on the sub-block
in the Negative Wrapped Convolution that uses the NTT and pre-process
modules, $\text{NTT}(\tilde{f})\,\Circle\,\text{NTT}(\tilde{g})$
where $\tilde{f}=\text{pre-process}(f)$ and $\tilde{g}=\text{pre-process}(g)$.
We have devised an error detection scheme at the algorithm level on
the component-wise NTT multiplication sub-block, $\text{NTT}(\tilde{f})\,\Circle\,\text{NTT}(\tilde{g})$,
as well as an error detection scheme for the pre-process function,
involving careful utilization of recomputation using shifted operands.
This error detection scheme can be applied to designs which require
polynomial multiplication using the NTT. We used Negative Wrapped
Convolution in our proposed error detection scheme. 

\subsubsection{Component-wise NTT Multiplication}

Our focus is on $\text{NTT}(\tilde{f})\,\Circle\,\text{NTT}(\tilde{g})$
operation in this section. Jou \emph{et al.} \citep{jou1988fault}
proposed an algorithm level error detection scheme on FFT network.
Their proposed error detection scheme is depicted in Fig. 1. In order
to achieve an efficient scheme, we have adapted the aforementioned
approach to the component-wise NTT multiplication. The primary focus
is to use an error detection scheme with small overhead and high error
detection ratio with respect to efficient realizations of the NTT.
From the NTT definition, we can use the relations presented below:
\begin{figure}[t]
\centering{}\includegraphics[clip,width=0.58\columnwidth,viewport=0bp 0bp 960bp 540bp]{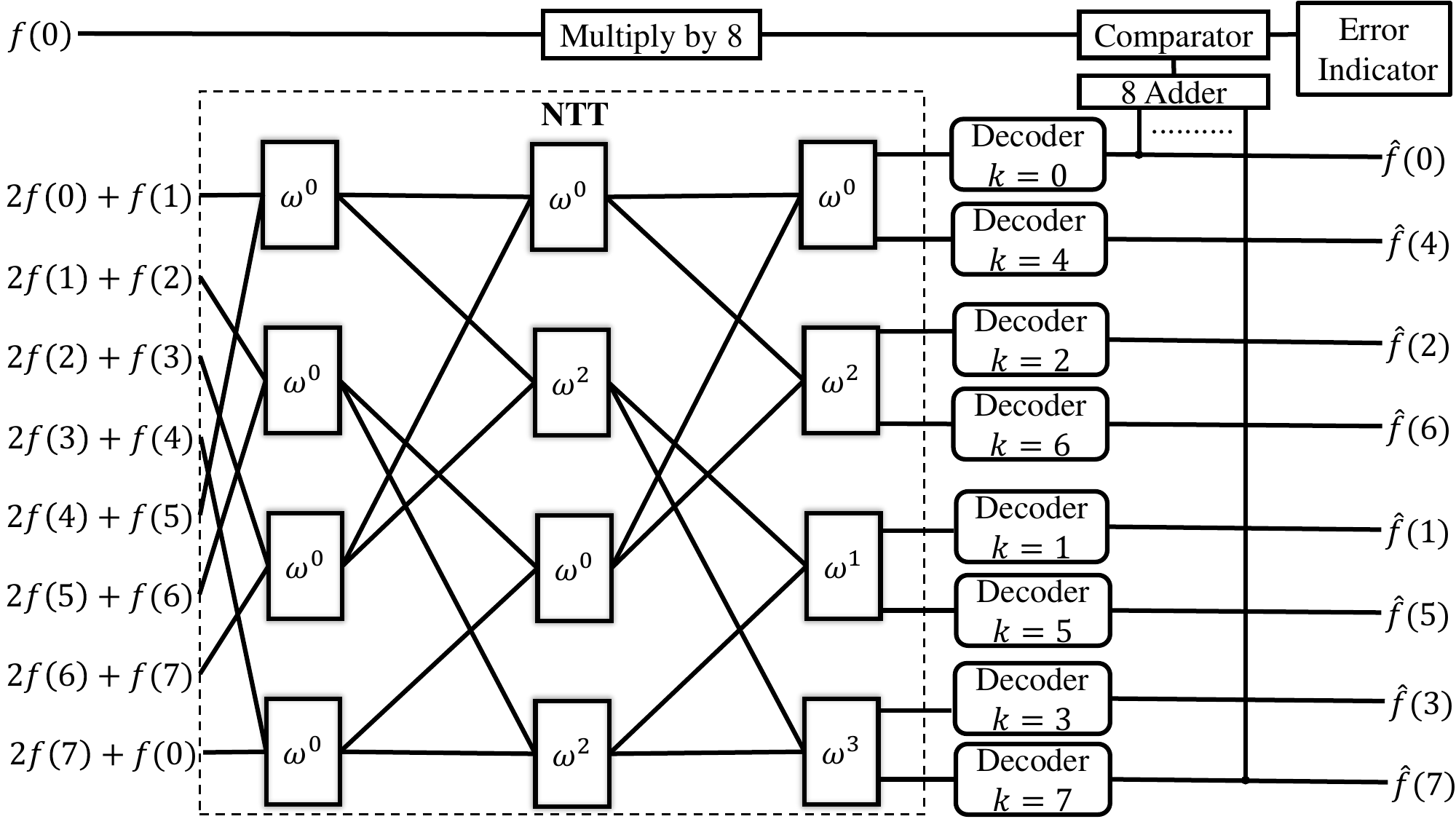}\caption{Concurrent error detection scheme for the NTT operation.}
\end{figure}

\begin{equation}
\hat{f}(0)=\sum_{j=0}^{n-1}\tilde{f}(j)\omega^{0}=\sum_{j=0}^{n-1}\tilde{f}(j),\quad j=0,1,...,n-1,
\end{equation}
\begin{equation}
\hat{g}(0)=\sum_{j=0}^{n-1}\tilde{g}(j)\omega^{0}=\sum_{j=0}^{n-1}\tilde{g}(j),\quad j=0,1,...,n-1.
\end{equation}
The result of the NTT multiplication for index 0 can be written as:

\begin{equation}
\hat{f}(0)\hat{g}(0)=\sum_{j=0}^{n-1}\tilde{f}(j)\sum_{j=0}^{n-1}\tilde{g}(j).
\end{equation}

We note that (7) is used in our proposed error detection scheme. If
we rotate the input sequence of (1) by one, then we get:

$\left[\begin{array}{c}
\hat{f}(0)\\
\hat{f}(1)\\
\hat{f}(2)\\
.\\
\hat{f}(N-1)
\end{array}\right]=\rho*\left[\begin{array}{c}
\tilde{f}(1)\\
\tilde{f}(2)\\
.\\
\tilde{f}(N-1)\\
\tilde{f}(0)
\end{array}\right]$or $\left[\begin{array}{c}
\hat{f}(0)/\omega^{0}\\
\hat{f}(1)/\omega^{1}\\
\hat{f}(2)/\omega^{2}\\
.\\
\hat{f}(n-1)/\omega^{(n-1)}
\end{array}\right]=\theta*\left[\begin{array}{c}
\tilde{f}(1)\\
\tilde{f}(2)\\
.\\
\tilde{f}(n-1)\\
\tilde{f}(0)
\end{array}\right]$ where the symbol \textasteriskcentered{} denotes matrix multiplication,
$\rho=\left[\begin{array}{cccc}
\omega^{0} & \omega^{0} & ... & \omega^{0}\\
\omega^{1} & \omega^{2} & ... & \omega^{0}\\
\omega^{2} & \omega^{4} & ... & \omega^{0}\\
. & . & ... & .\\
\omega^{n-1} & \omega^{2(n-1)} & ... & \omega^{0}
\end{array}\right]$ and $\theta=$$\left[\begin{array}{cccc}
\omega^{0} & \omega^{0} & ... & \omega^{0}\\
\omega^{0} & \omega^{1} & ... & \omega^{n-1}\\
\omega^{0} & \omega^{2} & ... & \omega^{2(n-1)}\\
. & . & ... & .\\
\omega^{0} & \omega^{n-1} & ... & \omega^{(n-1)^{2}}
\end{array}\right]$.

From $\theta*\left[\begin{array}{c}
\alpha\tilde{f}(0)\\
\alpha\tilde{f}(1)\\
\alpha\tilde{f}(2)\\
.\\
\tilde{\alpha f}(n-1)
\end{array}\right]$$\,+\,\theta*\left[\begin{array}{c}
\tilde{\beta f}(1)\\
\beta\tilde{f}(2)\\
.\\
\tilde{\beta f}(n-1)\\
\tilde{\beta f}(0)
\end{array}\right]$ , we get the following: 

\begin{align}
\left[\begin{array}{c}
\hat{f}(0)(\alpha+\beta\omega^{-0})\\
\hat{f}(1)(\alpha+\beta\omega^{-1})\\
\hat{f}(2)(\alpha+\beta\omega^{-2})\\
.\\
\hat{f}(n-1)(\alpha+\beta\omega^{-(n-1)})
\end{array}\right] & =\theta*\left[\begin{array}{c}
\alpha\tilde{f}(0)+\beta\tilde{f}(1)\\
\alpha\tilde{f}(1)+\beta\tilde{f}(2)\\
.\\
\alpha\tilde{f}(n-2)+\beta\tilde{f}(n-1)\\
\alpha\tilde{f}(n-1)+\beta\tilde{f}(0)
\end{array}\right],
\end{align}
where $\alpha$ and $\beta$ in (8) are scalars which can be selected
arbitrarily.  Let us denote $\tilde{f}^{m}$ and $\tilde{g}^{m}$,
respectively, as the inputs which are shifted by $m$$.$ In other
words, we can achieve:

\begin{equation}
\hat{f}=\frac{1}{(\alpha+\beta\omega^{-k})}NTT(\alpha\tilde{f}+\beta\tilde{f}^{1}),
\end{equation}

\begin{equation}
\hat{g}=\frac{1}{(\alpha+\beta\omega^{-k})}NTT(\alpha\tilde{g}+\beta\tilde{g}^{1}).
\end{equation}
Let us denote $h$ as the component-wise NTT multiplication of $\tilde{f}$
and $\tilde{g}$$.$ We can get:

\begin{align}
h & =NTT(\tilde{f})\,\Circle\,NTT(\tilde{g})=\frac{1}{(\alpha+\beta\omega^{-k})^{2}}NTT(\alpha\tilde{f}+\beta\tilde{f}^{1})\,\Circle\,NTT(\alpha\tilde{g}+\beta\tilde{g}^{1}).
\end{align}
The proposed error detection scheme is depicted in Fig. 2. The Decoder\_2
sub-block multiplies its input with $\frac{1}{(\alpha+\beta\omega^{-k})^{2}}$,
which is the inverse of $(\alpha+\beta\omega^{-k})^{2}$ in $\mathbb{R}=\mathbb{Z}[X]/(X^{n}+1)$.
The scheme for error detection involves comparing $h(0)$, obtained
from (11), with $\sum_{j=0}^{n-1}\tilde{f}(j)\sum_{j=0}^{n-1}\tilde{g}(j)$
obtained from (7).
\begin{figure}[t]
\centering{}\includegraphics[clip,width=0.6\columnwidth,viewport=0bp 0bp 960bp 540bp]{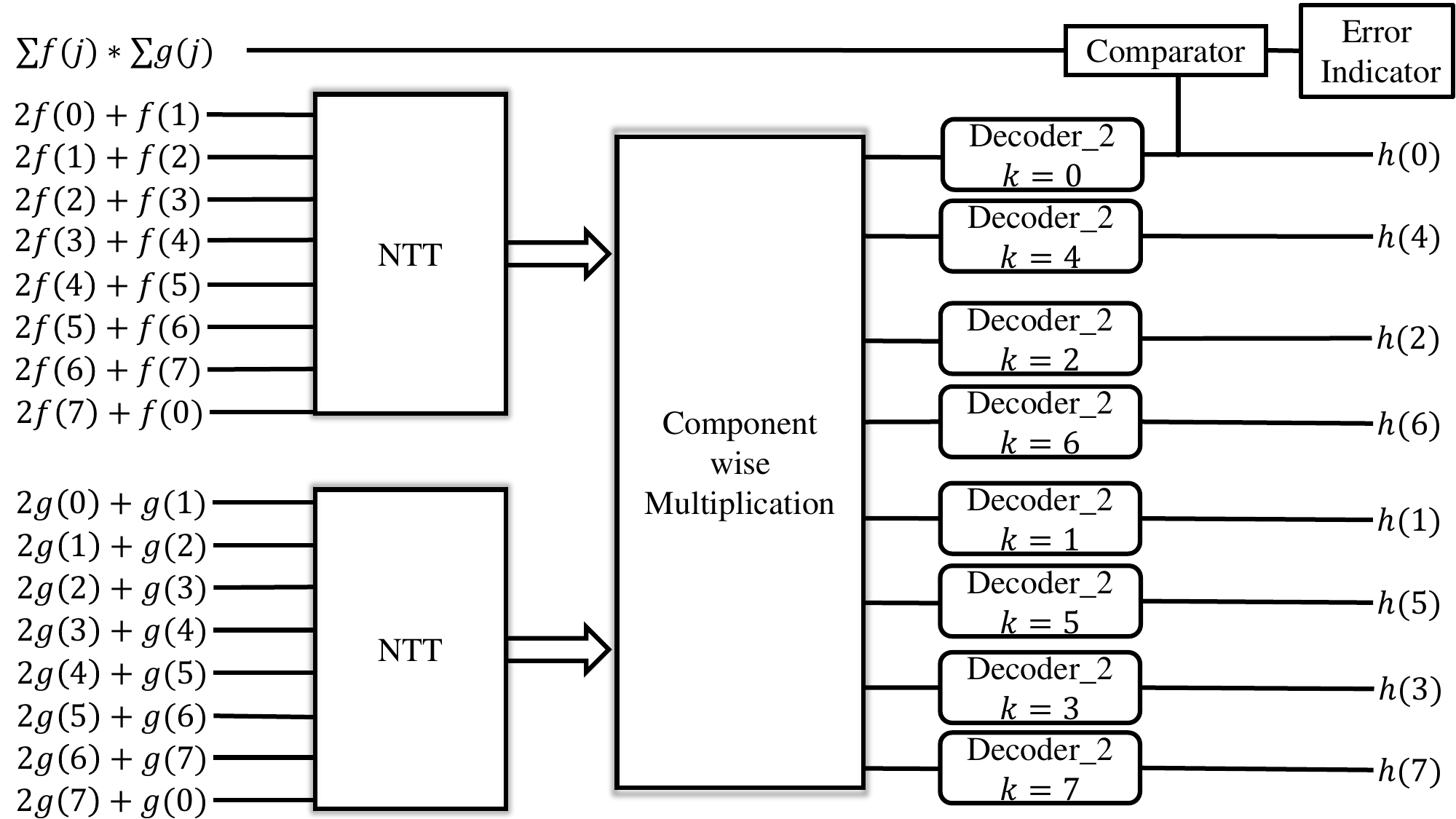}\caption{Proposed algorithm level error detection scheme for the NTT multiplication
module using Negative Wrapped Convolution.}
\end{figure}

\subsubsection{Pre-processor}

The pre-processor module performs element-by-element multiplication
on two input arrays. In proposing the error detection scheme, we have
applied an additional recomputing with shifted operands. As shown
in Fig. 3, the encoding and decoding modules constitute shifting which
is free in hardware. The pre-processing module considers a slight
overhead linked to the NTT multiplication. As a result, the error
detection scheme employing recomputation does not notably impact the
overall performance.

\subsection{Kyber}

To apply the Negative Wrapped Convolution method, it is necessary
that $2N\mid(q-1)$. For the Kyber parameters with the prime $q=3329$
and $N=256$, the base field $\mathbb{Z}_{q}$ includes 256-th roots
of unity but lacks 512-th roots. Therefore, Negative Wrapped Convolution
cannot be used in Kyber. We proposed below our error detection scheme
on the Kyber NTT module (equations (2) and (3)).

\subsubsection{NTT Module}

By expressing equations (2) and (3) as a matrix multiplication, we
obtain the following relation where $N=256$ with respect to Kyber
parameters: 

$f_{even}^{'}=\left[\begin{array}{c}
\hat{f}(0)\\
\hat{f}(2)\\
.\\
\hat{f}(252)\\
\hat{f}(N-2)
\end{array}\right]=\partial*f_{even}$, and $f_{odd}^{'}=\left[\begin{array}{c}
\hat{f}(1)\\
\hat{f}(3)\\
.\\
\hat{f}(253)\\
\hat{f}(N-1)
\end{array}\right]=\partial*f_{odd}$, where $f_{even}=\left[\begin{array}{c}
f(0)\\
f(2)\\
.\\
f(252))\\
f(N-2)
\end{array}\right]$, 

$f_{odd}=\left[\begin{array}{c}
f(1)\\
f(3)\\
.\\
f(253)\\
f(N-1)
\end{array}\right]$, and $\partial=\left[\begin{array}{ccccc}
1 & \omega & \omega^{2} & ... & \omega^{127}\\
1 & \omega^{2b(1)+1} & \omega^{2(2b(1)+1)} & ... & \omega^{127(2b(1)+1)}\\
1 & \omega^{2b(2)+1} & \omega^{2(2b(2)+1)} & ... & \omega^{127(2b(2)+1)}\\
. & . & . & ... & .\\
1 & \omega^{2b(127)+1} & \omega^{2(2b(127)+1)} & ... & \omega^{127(2b(127)+1)}
\end{array}\right]$.

\begin{figure}[t]
\centering{}\vspace{-2cm}

\includegraphics[clip,width=0.65\columnwidth,viewport=0bp 0bp 960bp 540bp]{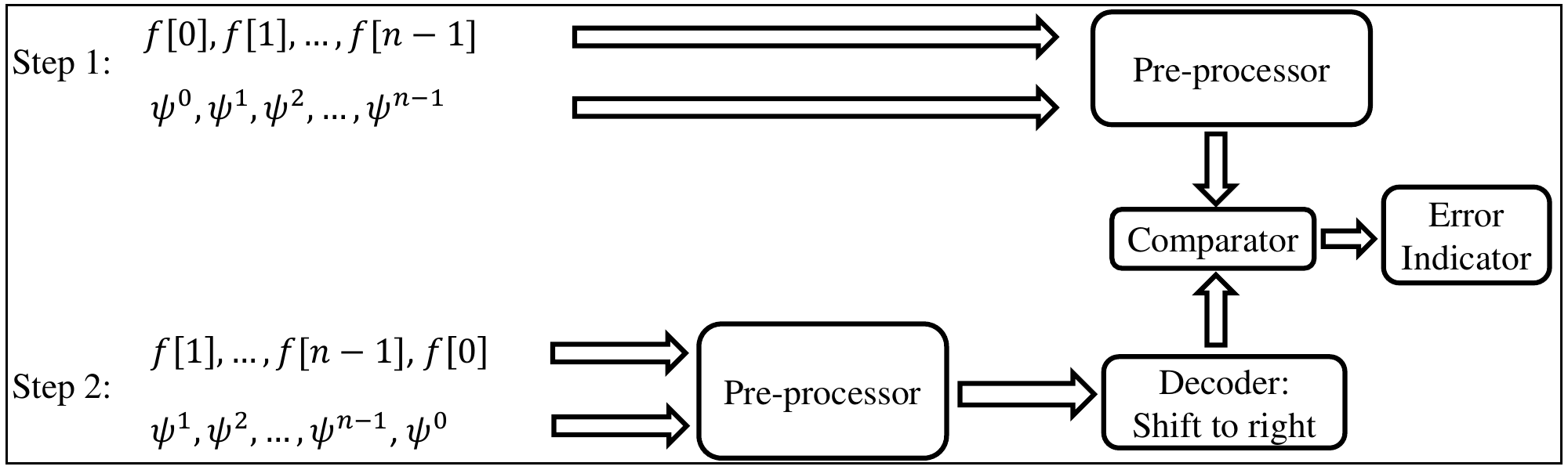}

\caption{Proposed error detection scheme for pre-process module through recompuation
with shifted operands.}
\end{figure}
If we rotate the input sequence of $f_{even}^{'}$ and $f_{odd}^{'}$
by one, by using the symmetric and periodicity properties ($\omega^{k+N/2}=-\omega^{k}$
and $\omega^{k+N}=\omega^{k})$ of NTT, we get: 

\begin{equation}
f_{even}^{''}=\left[\begin{array}{c}
\frac{\hat{f}(0)-2f(0)}{\omega^{2b(0)+1}}\\
\frac{\hat{f}(2)-2f(0)}{\omega^{2b(1)+1}}\\
.\\
\frac{\hat{f}(252)-2f(0)}{\omega^{2b(126)+1}}\\
\frac{\hat{f}(N-2)-2f(0)}{\omega^{2b(\frac{N-2}{2})+1}}
\end{array}\right]=\partial*\left[\begin{array}{c}
f(2)\\
f(4)\\
.\\
f(N-2)\\
f(0)
\end{array}\right],
\end{equation}

\begin{equation}
f_{odd}^{''}=\left[\begin{array}{c}
\frac{\hat{f}(1)-2f(1)}{\omega^{2b(0)+1}}\\
\frac{\hat{f}(3)-2f(1)}{\omega^{2b(1)+1}}\\
.\\
\frac{\hat{f}(253)-2f(1)}{\omega^{2b(126)+1}}\\
\frac{\hat{f}(N-1)-2f(1)}{\omega^{2b(\frac{N-2}{2})+1}}
\end{array}\right]=\partial*\left[\begin{array}{c}
f(3)\\
f(5)\\
.\\
f(N-1)\\
f(1)
\end{array}\right].
\end{equation}

From $\alpha$$f_{even}^{'}$ + $\beta$$f_{even}^{''}$ and $\alpha$$f_{odd}^{'}$
+ $\beta$$f_{odd}^{''}$ , we get the following:
\begin{align}
\left[\begin{array}{c}
\alpha\hat{f}(0)+\frac{\beta(\hat{f}(0)-2f(0))}{\omega^{2b(0)+1}}\\
\alpha\hat{f}(2)+\frac{\beta(\hat{f}(2)-2f(0))}{\omega^{2b(1)+1}}\\
.\\
\alpha\hat{f}(252)+\frac{\beta(\hat{f}(252)-2f(0))}{\omega^{2b(126)+1}}\\
\alpha\hat{f}(N-2)+\frac{\beta(\hat{f}(N-2)-2f(0))}{\omega^{2b(127)+1}}
\end{array}\right] & =\partial*\left[\begin{array}{c}
\alpha f(0)+\beta f(2)\\
\alpha f(2)+\beta f(4)\\
.\\
\alpha f(252)+\beta f(N-2)\\
\alpha f(N-2)+\beta f(0)
\end{array}\right],
\end{align}

\begin{align}
\left[\begin{array}{c}
\alpha\hat{f}(1)+\frac{\beta(\hat{f}(1)-2f(1))}{\omega^{2b(0)+1}}\\
\alpha\hat{f}(3)+\frac{\beta(\hat{f}(3)-2f(1))}{\omega^{2b(1)+1}}\\
.\\
\alpha\hat{f}(253)+\frac{\beta(\hat{f}(253)-2f(1))}{\omega^{2b(126)+1}}\\
\alpha\hat{f}(N-1)+\frac{\beta(\hat{f}(N-1)-2f(1))}{\omega^{2b(127)+1}}
\end{array}\right] & =\partial*\left[\begin{array}{c}
\alpha f(1)+\beta f(3)\\
\alpha f(3)+\beta f(5)\\
.\\
\alpha f(253)+\beta f(N-1)\\
\alpha f(N-1)+\beta f(1)
\end{array}\right].
\end{align}

Let us denote $f_{even}^{1}$ and $f_{odd}^{1}$ , respectively, as
the $f_{even}$ and $f_{odd}$ that are shifted by $1$. If we call
equations (2) and (3) NTT\_Kyber, we can achieve:
\begin{equation}
\hat{f}(2k)=\text{\ensuremath{\frac{1}{(\frac{\beta}{\omega^{2b(k)+1}}+\alpha)}} (NTT\_Kyber}(\alpha f(2k)+\beta f(2k)^{1})+\frac{2\beta f(0)}{\omega^{2b(k)+1}}),\quad k=0,1,...,127,
\end{equation}

\begin{equation}
\hat{f}(2k+1)=\text{\ensuremath{\frac{1}{(\frac{\beta}{\omega^{2b(k)+1}}+\alpha)}} (NTT\_Kyber}(\alpha f(2k+1)+\beta f(2k+1)^{1})+\frac{2\beta f(1)}{\omega^{2b(k)+1}}),\quad k=0,1,...,127.
\end{equation}

We can use the relation presented below for our error detection scheme:

\begin{equation}
128(f(0)+f(1))\text{ mod }q=\sum_{j=0}^{n-1}\hat{f}(j)\text{ mod \ensuremath{q},\ensuremath{\quad q=3329}}.
\end{equation}

The proposed error detection scheme is depicted in Fig. 4. The Decoder\_3
sub-block multiplies its input with $\frac{1}{(\frac{\beta}{\omega^{2b(k)+1}}+\alpha)}$,
which is the inverse of $(\frac{\beta}{\omega^{2b(k)+1}}+\alpha)$
in $\mathbb{R}=\mathbb{Z}[X]/(X^{n}+1)$. The scheme for error detection
involves comparing $128(f(0)+f(1))$ mod $q$, with $\sum_{j=0}^{n-1}\hat{f}(j)$
mod $q$.

\begin{figure}[t]
\centering{}\includegraphics[clip,width=0.6\columnwidth,viewport=0bp 0bp 960bp 540bp]{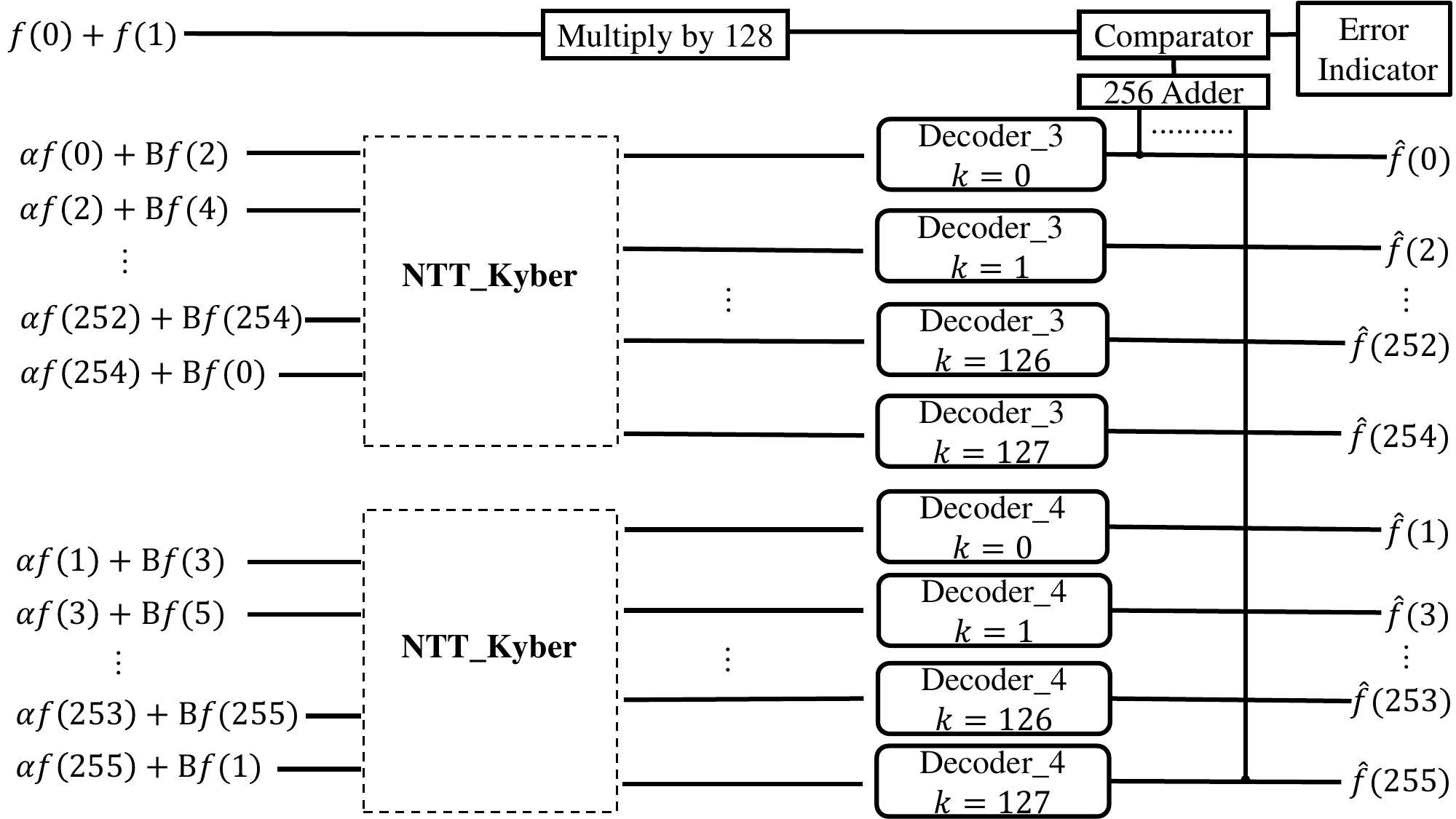}\caption{Proposed algorithm level error detection scheme for the NTT utilized
in Kyber.}
\end{figure}

\section{Error Coverage Simulation Results}

\subsection{Fault Model}

The butterfly module is a key element in the NTT that enables efficient
modular transformations, much like its function in the FFT. It processes
two input elements by calculating their sum and difference. Its ability
to handle modular arithmetic and root of unity multiplications is
essential for enhancing the performance of the NTT.

As depicted in Fig. 5, faults can happen at modules \{1,2, or 3\}
marked in the figure, which may result in erroneous outputs of a butterfly
sub-block used in the NTT module. Fault occurrence follows a normal
distribution within butterfly modules in Fig. 2 and Fig. 4. Furthermore,
as we provide fault detection schemes over the NTT multiplication,
faults can happen during component wise multiplication module. The
NTT process involves numerous butterfly operations, where faults may
occur at any of these operations. Within each butterfly, faults can
happen at positions \{1, 2, or 3\}. In our fault model, no two faults
are allowed to occur in the same butterfly operation. In the normal
mode fault injection method, if the NTT process contains $N$ butterfly
operations and $F$ represents the total number of faults, each butterfly
operation has a probability of $\frac{F}{N}$ of being faulty. Moreover,
if a butterfly is faulty, the probability of the fault occurring at
position \{1, 2, or 3\} is $\frac{1}{3}$. 

In the burst mode fault injection method, once a faulty butterfly
operation is randomly chosen, all subsequent butterfly operations
will also be faulty.
\begin{figure}[t]
\centering{}

\vspace{-2.1cm}
\includegraphics[clip,width=0.6\columnwidth,viewport=0bp 0bp 960bp 540bp]{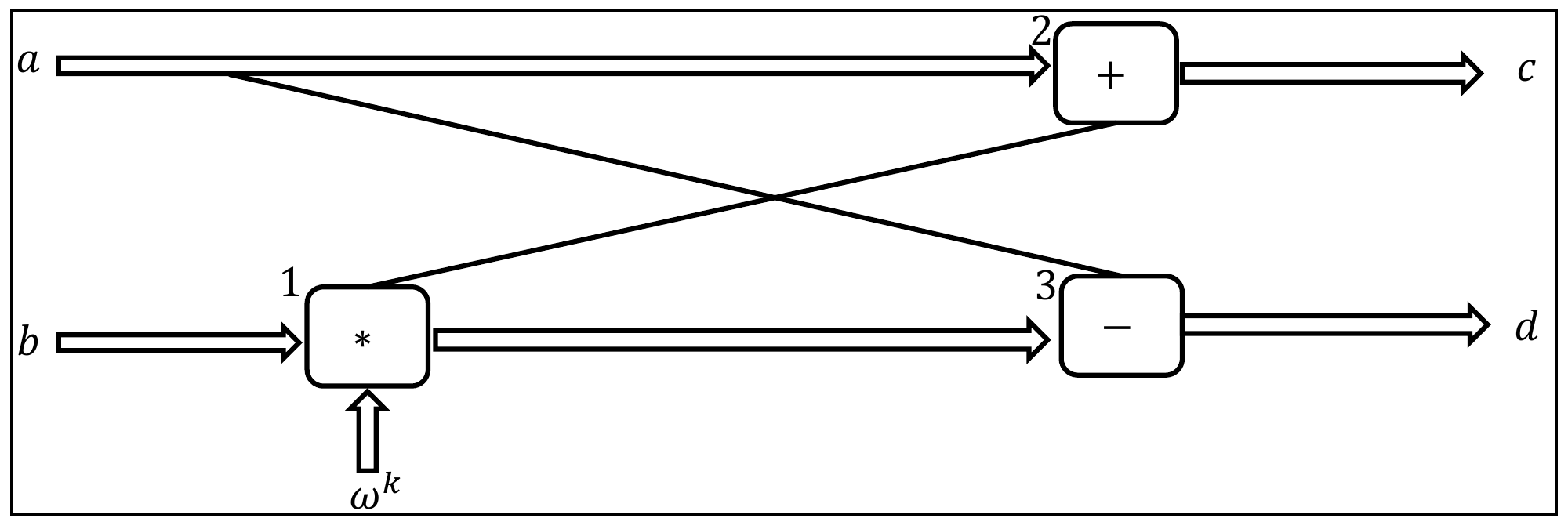}

\caption{The utilized fault model in this work for the butterfly sub-block.}
\end{figure}

\subsection{Fault Simulation}

To assess the error coverage, we employed Python3 to perform simulations,
applying the described fault models and error detection techniques
to both the NTT and pre-process modules. 

\subsubsection{Negative Wrapped Convolution}

The simulations encompassed a million samples, using parameters identical
to the standard Kyber Round 1. As shown in Table 1, with increase
in the number of faults, we can attain an error detection ratio close
to 100\% for both modules. Higher error coverage is achievable for
identifying burst errors. 

The table provides data comparing the effectiveness of the proposed
fault injection on two cryptographic components: the Pre-process stage
and the NTT Multiplication stage, based on the number of faults injected
(1, 2, 4, 8, or 16). For the Pre-process component, for a single fault,
we already get high error coverage of 99.7\%. As the number of injected
faults increases, the ratio quickly reaches 100\%. In contrast, the
NTT Multiplication stage is more difficult to protect, but eventually
we get to close to 100\%. With one injected fault, 53\% is the detection
ratio. However, the success rate increases as more faults are injected:
70.6\% for 2 faults, 90.9\% for 4, 99\% for 8, and 99.9\% for 16 faults.
\begin{table}

\caption{Negative Wrapped Convolution achieved error detection ratios of the
proposed schemes\- with 1, 2, 4, 8, and 16 faults occurrence with
normal fault injection method for one million samples}

\centering

\begin{tabular}{c|c|cc}
\multirow{2}{*}{\textbf{Parameters}} & \multirow{2}{*}{} & \multicolumn{2}{c}{$^{1}n=256,^{2}\omega=3,844$,}\tabularnewline
 &  & \multicolumn{2}{c}{$^{3}q=7,681$}\tabularnewline
\hline 
\multirow{2}{*}{\textbf{Component}} & \multirow{2}{*}{} & \multirow{2}{*}{\textbf{Pre-process}} & \textbf{NTT}\tabularnewline
 &  &  & \textbf{Multiplication}\tabularnewline
\hline 
 & 1 & 99.7\% & 53\%\tabularnewline
\cline{2-4} \cline{3-4} \cline{4-4} 
\textbf{Number of} & 2 & 99.9\% & 70.6\%\tabularnewline
\cline{2-4} \cline{3-4} \cline{4-4} 
\textbf{faults} & 4 & 100\% & 90.9\%\tabularnewline
\cline{2-4} \cline{3-4} \cline{4-4} 
 & 8 & 100\% & 99\%\tabularnewline
\cline{2-4} \cline{3-4} \cline{4-4} 
 & 16 & 100\% & 99.9\%\tabularnewline
\hline 
\end{tabular}\vspace{0.2cm}

\raggedright{}$^{1}$$n=$ Polynomial size, $^{2}$$\omega=$ Twiddle
factor, $^{3}$$q=$ Prime number
\end{table}

\subsubsection{Kyber}

In the standard Kyber Round 3 NTT implementation, there are 896 butterfly
operations, corresponding to the polynomial size of 256. A fault can
potentially occur at any of these 896 butterfly operations. The simulations
involved one million samples, using the same parameters as in Kyber
Round 3. 

As indicated in Table 2, increasing the number of faults leads to
an error detection ratio approaching 100\% in both normal and burst
modes. In Kyber's NTT implementation, the error detection ratios for
various fault occurrences were analyzed using one million samples,
with parameters set to $n=256$, $\omega=17,$ and $q=3329$. The
results show that as the number of injected faults increases, the
error detection ratio improves significantly. With just 1 fault, 74.9\%
of the errors were detected. This detection rate jumps to 93.45\%
with 2 faults. With 4 faults, the detection ratio reaches 99.49\%,
and by injecting 8 faults, it further increases to 99.95\%. Finally,
with 16 faults, the error detection ratio hits 100\%, indicating complete
detection of faults. This trend highlights the robustness of the error
detection mechanism in Kyber\textquoteright s NTT when subjected to
increasing numbers of faults. A higher number of faults leads to near-perfect
or perfect error detection, suggesting the scheme becomes more effective
at fault identification as fault intensity increases. 

The NTT in Kyber was also tested using a burst fault model, with error
detection ratios evaluated over one million samples. In the burst
fault model, multiple consecutive bits in the data are corrupted simultaneously.
The number of burst faults was varied, and the error detection ratios
achieved by the proposed fault detection scheme are as follows:
\begin{itemize}
\item With 2 burst faults, the scheme detected errors with a success rate
of \textbf{93.82\%}. 
\item When 3 burst faults occurred, the detection rate increased to \textbf{98.3\%}. 
\item With 4 burst faults, the detection ratio improved further to \textbf{99.42\%}. 
\item At 5 burst faults, the scheme achieved a near-perfect detection rate
of \textbf{99.76\%}.
\item For 6 burst faults, the detection rate was nearly flawless, reaching
\textbf{99.8\%}.
\end{itemize}
The trend shows that as the number of burst faults increases, the
error detection capability of the scheme improves, approaching almost
100\% accuracy with 5 or more faults. This indicates the robustness
of the detection method, particularly when dealing with higher levels
of fault occurrences.
\begin{table}
\caption{NTT in Kyber achieved error detection ratios of the proposed schemes
with normal and burst fault injection methods for one million samples}

\centering

\begin{tabular}{c|c|c|c|c}
\multirow{2}{*}{\textbf{Parameters}} &  & \multicolumn{3}{c}{$n=256,\omega=17$}\tabularnewline
 &  & \multicolumn{3}{c}{$q=3,329$}\tabularnewline
\hline 
\textbf{Fault injection method} &  & Normal mode & \multicolumn{2}{c}{Burst mode}\tabularnewline
\hline 
 & 1 & 74.9\% & 2 & 93.82\%\tabularnewline
\cline{2-5} \cline{3-5} \cline{4-5} \cline{5-5} 
\textbf{Number of} & 2 & 93.45\% & 3 & 98.3\%\tabularnewline
\cline{2-5} \cline{3-5} \cline{4-5} \cline{5-5} 
\textbf{faults} & 4 & 99.49\% & 4 & 99.42\%\tabularnewline
\cline{2-5} \cline{3-5} \cline{4-5} \cline{5-5} 
 & 8 & 99.95\% & 5 & 99.76\%\tabularnewline
\cline{2-5} \cline{3-5} \cline{4-5} \cline{5-5} 
 & 16 & 100\% & 6 & 99.8\%\tabularnewline
\hline 
\end{tabular}
\end{table}

\section{Xilinx/AMD FPGA and Software Implementation results}

\subsection{Negative Wrapped Convolution}

To confirm the efficiency of our presented approach, we chose to assess
its performance by applying it to the NTT multiplication involving
a 256-degree polynomial. We conducted a benchmark for implementation
on different FPGAs, i.e., Xilinx/AMD Zynq Ultrascale+ and Artix-7.
The Xilinx/AMD Zynq Ultrascale+ is a high-performance, multi-core
SoC (System on Chip) that integrates programmable logic with ARM Cortex-A53
processors. It is designed for complex applications requiring significant
processing power, such as AI, 5G, and automotive systems. It offers
advanced features like multi-core processing, high-speed I/O, and
power optimization.

In contrast, the Xilinx Artix-7 is a lower-cost, mid-range FPGA aimed
at applications that need efficient power usage and moderate performance,
such as communications, video processing, and embedded systems. It
does not include integrated processors, relying solely on programmable
logic.

Key differences include the Zynq Ultrascale+'s integrated ARM processors
and higher performance capabilities, whereas the Artix-7 focuses on
power efficiency and simpler, more cost-effective designs.

The results clearly demonstrate that our proposed efficient error
detection schemes maintain a modest overhead while effectively achieving
a high level of fault detection. We utilized the High-Level Synthesis
(HLS) Vitis development environment of AMD/Xilinx to transform our
proposed schemes into Register Transfer Level (RTL) hardware descriptions.
The generated IP imported to Vivado to report power, utilization,
and latency. Tables 3 and 4 demonstrate the results of our implementations
and the derivations for area, timing, power, and energy. In the Vivado
synthesis tool context, area is determined by combining Slices and
DSPs, with a conversion ratio where one DSP is deemed equivalent to
100 Slices. Let us first go over Table 4. This table compares the
implementation of two designs (our scheme vs. baseline) on the AMD/Xilinx
Zynq Ultrascale+ FPGA (xczu4ev-sfvc784-2-i) in terms of area, power,
and performance. Our scheme, which includes error detection, uses
slightly more hardware resources than the baseline, with higher values
for LUTs (1,469 vs. 1,435), FFs (1,188 vs. 1,122), CLBs (343 vs. 340),
and DSPs (33 vs. 30). Both designs have similar power consumption
(0.46W vs. 0.45W), but our scheme incurs higher latency (3,703 vs.
3,438 clock cycles), total time (26,439 ns vs. 24,547 ns), and energy
consumption (12,161 nJ vs. 11,046 nJ) due to the added complexity
of error detection. These are reasonable overheads for providing error
detection.

Moreover, as seen in Table 5, it summarizes the implementation results
of two designs (our scheme vs. baseline) on the AMD/Xilinx Artix-7
FPGA (xc7a75ti-ftg256-1L), comparing their area, power, and timing
performance. Our scheme, which includes error detection, uses more
resources than the baseline, with higher values for LUTs (1,569 vs.
1,530), FFs (1,693 vs. 1,597), and DSPs (33 vs. 30). Both designs
have similar power consumption, but our scheme consumes slightly more
(0.2W vs. 0.19W). However, our scheme has increased latency (3,749
vs. 3,482 clock cycles), total time (26,767 ns vs. 24,861 ns), and
energy consumption (5,353 nJ vs. 4,723 nJ), while providing fault
detection not found in the baseline design. Our proposed error detection
designs achieve a maximum operational frequency of around 140 MHz
across all the FPGAs. Our proposed error detection scheme imposes
a maximum overhead of 9\% in additional area and introduces a latency
increase of up to 13\% in clock cycles, at most considering different
designs. Our simulation and implementation code is publicly available
in our github\footnote{https://github.com/KasraAhmadi/NTT\_Error\_Detection}.

\subsubsection{Implementation Optimization}

Although HLS has shifted the design entry level of abstraction from
RTL to C/C++, practical implementation often requires significant
source code rewriting to make it HLS-ready. We have carefully taken
this into effort as discussed here. This includes the incorporation
of pragmas to attain satisfactory performance. Our intended area and
timing efforts implementation was realized by strategically inserting
the following pragmas into our program.

\begin{table}[t]
\caption{AMD/Xilinx Zynq Ultrascale+, xczu4ev-sfvc784-2-i Implementation Results}

\centering

\begin{tabular}{cccccc}
\toprule 
\textbf{Strategy} &  & \multicolumn{2}{c}{\textbf{Area}} & \multicolumn{2}{c}{\textbf{Timing}}\tabularnewline
 &  & \multicolumn{2}{c}{\textbf{Effort}} & \multicolumn{2}{c}{\textbf{Effort}}\tabularnewline
\midrule 
\multirow{2}{*}{\textbf{Scheme}} & \multirow{2}{*}{} & Our & Baseline & Our & Baseline\tabularnewline
 &  & scheme & work & scheme & work\tabularnewline
\midrule 
 & \multicolumn{1}{c}{LUTs} & 1,469 & 1,435 & 3,207 & 3,165\tabularnewline
\cmidrule{2-6} \cmidrule{3-6} \cmidrule{4-6} \cmidrule{5-6} \cmidrule{6-6} 
\multirow{2}{*}{\textbf{Area}} & \multicolumn{1}{c}{FFs} & 1,188 & 1,122 & 2,227 & 2,161\tabularnewline
\cmidrule{2-6} \cmidrule{3-6} \cmidrule{4-6} \cmidrule{5-6} \cmidrule{6-6} 
 & \multicolumn{1}{c}{CLBs} & 343 & 340 & 733 & 683\tabularnewline
\cmidrule{2-6} \cmidrule{3-6} \cmidrule{4-6} \cmidrule{5-6} \cmidrule{6-6} 
 & \multicolumn{1}{c}{DSPs} & 33 & 30 & 54 & 51\tabularnewline
\midrule 
\textbf{Power (W)} &  & \multirow{2}{*}{0.46} & \multirow{2}{*}{0.45} & \multirow{2}{*}{0.52} & \multirow{2}{*}{0.51}\tabularnewline
\textbf{@ 140 MHz} &  &  &  &  & \tabularnewline
\midrule 
 & Latency & \multirow{2}{*}{3,703} & \multirow{2}{*}{3,438} & \multirow{2}{*}{2,241} & \multirow{2}{*}{1,979}\tabularnewline
\multirow{2}{*}{\textbf{Timing}} & {[}CCs{]} &  &  &  & \tabularnewline
\cmidrule{2-6} \cmidrule{3-6} \cmidrule{4-6} \cmidrule{5-6} \cmidrule{6-6} 
 & \multicolumn{1}{c}{Total time} & \multirow{2}{*}{26,439} & \multirow{2}{*}{24,547} & \multirow{2}{*}{16,000} & \multirow{2}{*}{14,130}\tabularnewline
 & {[}ns{]} &  &  &  & \tabularnewline
\midrule 
\textbf{Energy (nJ)} & \multicolumn{1}{c}{} & 12,161 & 11,046 & 8,320 & 7,206\tabularnewline
\bottomrule
\end{tabular}

\vspace{0.2cm}
 \raggedright{}Our Scheme: The design which includes error detection
scheme.

Baseline work: The design which does not include any error detection
scheme.
\end{table}

\begin{table}[t]
\caption{AMD/Xilinx Artix-7, xc7a75ti-ftg256-1L Implementation Results}

\centering

\begin{tabular}{cccccc}
\toprule 
\textbf{Strategy} &  & \multicolumn{2}{c}{\textbf{Area}} & \multicolumn{2}{c}{\textbf{Timing}}\tabularnewline
 &  & \multicolumn{2}{c}{\textbf{Effort}} & \multicolumn{2}{c}{\textbf{Effort}}\tabularnewline
\midrule 
\multirow{2}{*}{\textbf{Scheme}} & \multirow{2}{*}{} & Our & Baseline & Our & Baseline\tabularnewline
 &  & scheme & work & scheme & work\tabularnewline
\midrule 
 & \multicolumn{1}{c}{LUTs} & 1,569 & 1,530 & 2,956 & 2,939\tabularnewline
\cmidrule{2-6} \cmidrule{3-6} \cmidrule{4-6} \cmidrule{5-6} \cmidrule{6-6} 
\multirow{2}{*}{\textbf{Area}} & \multicolumn{1}{c}{FFs} & 1,693 & 1,597 & 2,979 & 2,883\tabularnewline
\cmidrule{2-6} \cmidrule{3-6} \cmidrule{4-6} \cmidrule{5-6} \cmidrule{6-6} 
 & \multicolumn{1}{c}{SLICEs} & 673 & 656 & 1,240 & 1,253\tabularnewline
\cmidrule{2-6} \cmidrule{3-6} \cmidrule{4-6} \cmidrule{5-6} \cmidrule{6-6} 
 & \multicolumn{1}{c}{DSPs} & 33 & 30 & 54 & 51\tabularnewline
\midrule 
\textbf{Power (W)} &  & \multirow{2}{*}{0.2} & \multirow{2}{*}{0.19} & \multirow{2}{*}{0.28} & \multirow{2}{*}{0.26}\tabularnewline
\textbf{@ 140 MHz} &  &  &  &  & \tabularnewline
\midrule 
 & Latency & \multirow{2}{*}{3,749} & \multirow{2}{*}{3,482} & \multirow{2}{*}{2,267} & \multirow{2}{*}{2,003}\tabularnewline
\multirow{2}{*}{\textbf{Timing}} & {[}CCs{]} &  &  &  & \tabularnewline
\cmidrule{2-6} \cmidrule{3-6} \cmidrule{4-6} \cmidrule{5-6} \cmidrule{6-6} 
 & \multicolumn{1}{c}{Total time} & \multirow{2}{*}{26,767} & \multirow{2}{*}{24,861} & \multirow{2}{*}{16,186} & \multirow{2}{*}{14,301}\tabularnewline
 & {[}ns{]} &  &  &  & \tabularnewline
\midrule 
\textbf{Energy (nJ)} & \multicolumn{1}{c}{} & 5,353 & 4,723 & 4,532 & 3,718\tabularnewline
\bottomrule
\end{tabular}
\end{table}

\subsubsection{Pre-calculate the decoder module values}

To enhance the efficiency of the proposed error detection scheme,
we have the option to pre-calculate the decoder module in (11), specifically
computing $\frac{1}{(\alpha+\beta\omega^{-k})^{2}}$ for various values
of the parameter $k$ and save them in memory. 

\subsubsection{Utilizing task-level pipelining}

We perform loop unrolling on the outer loop in Algorithm 1 (Line 2)
and transform the remaining code into a function named \textquotedbl stage,\textquotedbl{}
illustrating each stage in the NTT architecture. We undertook this
approach to leverage task-level pipelining, allowing functions and
loops to operate simultaneously. This enhances the concurrency of
the RTL implementation, resulting in an overall increase in design
throughput.

\subsubsection{Pipelining the stage function}

This pragma was inserted to facilitate instruction-level pipelining,
aiming to enhance throughput and clock frequency within each NTT stage
function. It is important to acknowledge that this optimization entails
the trade-off of utilizing extra digital resources.

\subsection{Kyber}

We implemented our proposed error detection scheme for the official
Kyber NIST submission (Round 3) on both x86-64 architecture and a
12th Gen Intel Core i7-12800H 2.4GHz processor, as well as on more
resource-constrained devices, such as the 2.4GHz quad-core ARM Cortex-A72
(Raspberry Pi). We utilized the PAPI library to evaluate the performance
of the software implementation and measure the overhead introduced
by our error detection scheme. The PAPI library provides a standardized
interface for accessing performance counters in CPUs, enabling researchers/developers
to measure and analyze the performance of applications efficiently.
Table 5 demonstrates the results of our implementation using standard
Kyber Round 3 implementation as baseline work on 2 different architectures.
The code for our simulation and implementation is accessible on our
GitHub. On the ARM Cortex-A72, the baseline work requires 8,859 clock
cycles to execute NTT operation in Kyber Round 3, whereas our scheme
takes 10,316 clock cycles, an increase of approximately 16.5\%. This
suggests that the additional error detection or enhancements come
with a performance cost in terms of clock cycles.

Similarly, on the Intel Core-i7, the baseline work completes in 5,814
clock cycles, while our scheme takes 7,443 clock cycles, an increase
of about 28\%. The performance overhead on the Intel Core-i7 is higher
compared to the ARM v8, indicating that the performance impact of
the error detection or other improvements is more significant on this
processor.

\begin{table}[t]
\caption{Software Implementation Results For the NTT in Kyber}

\begin{tabular}{ccc}
\hline 
 & \textbf{ARM v8 Cortex-A72 @ 1.5GHz} & \textbf{Intel Core-i7 @ 2.4GHz }\tabularnewline
\hline 
\textbf{Baseline work} & 8,859 clock cycles & 5,814 clock cycles\tabularnewline
\hline 
\textbf{Our Scheme} & 10,316 clock cycles & 7,443 clock cycles\tabularnewline
\hline 
\end{tabular}

\centering\vspace{0.2cm}

\raggedright{}Our Scheme: The design which includes error detection
scheme.

Baseline work: The design which does not include any error detection
scheme.
\end{table}

\subsubsection{Implementation Optimization}

According to equations (16) and (17), the overhead introduced by our
error detection scheme depends on the encoding and decoding modules.
The encoding module involves one addition and one shift operation.
For the decoder, we can store coefficients ($\frac{1}{(\frac{\beta}{\omega^{2b(k)+1}}+\alpha)}$
and $\frac{2\beta}{\omega^{2b(k)+1}}$) in memory to eliminate additional
computations. The decoding module comprises two multiplications and
one addition. 

\section{Conclusion}

In this paper, we proposed an efficient algorithm level error detection
scheme over polynomial multiplication using the NTT and Negative Wrapped
Convolution and the NTT operation in Kyber Round 3. We have aimed
at achieving low hardware overhead and low latency, suitable for deeply-embedded
systems. The presented scheme effectively safeguards cryptographic
algorithms that employ the NTT multiplication. By performing simulations,
we demonstrated that the proposed error detection scheme achieved
extensive coverage of faults. Moreover, we implemented our proposed
error detection schemes on two Xilinx/AMD FPGAs and two CPUs. Regarding
overhead and latency, the implementation led to minimal additional
expenses in hardware and software. With high error coverage and acceptable
overhead, the proposed schemes in this work are suitable for resource-constrained
and sensitive usage models.

\section*{Acknowledgments}

This work was supported by the US National Science Foundation (NSF)
through the award SaTC-1801488. 

\bibliographystyle{ACM-Reference-Format}
\bibliography{reference}

\end{document}